\documentclass[emulateapj]{emulateapj}

\slugcomment{LA-UR-06-1935}
\shorttitle{Ammonia in Gl 570D}
\shortauthors{Saumon et al.}

\def\wig#1{\mathrel{\hbox{\hbox to 0pt{%
          \lower.5ex\hbox{$\sim$}\hss}\raise.4ex\hbox{$#1$}}}}
\def\sss{\scriptscriptstyle}

\def\teff{T_{\rm eff}}
\def\sqr#1#2{{\vcenter{\vbox{\hrule height.#2pt 
  \hbox{\vrule width.#2pt height#1pt \kern#1pt 
  \vrule width.#2pt} 
  \hrule height.#2pt}}}}
\def\ref#1{\parindent=0pt\hangindent24pt\hangafter1
           \baselineskip=20pt{#1}\par }

\begin{document}

\title{Ammonia as a tracer of chemical equilibrium in the T7.5 dwarf Gliese 570D}

\author{D. Saumon\altaffilmark{1}, M. S. Marley\altaffilmark{2}, M. C. Cushing\altaffilmark{3,4}
        S. K. Leggett\altaffilmark{5}, T. L. Roellig\altaffilmark{6}, K. Lodders\altaffilmark{7}
        and R. S. Freedman\altaffilmark{8}}
\altaffiltext{1}{Los Alamos National Laboratory,  MS P365, Los Alamos, NM 87545; dsaumon@lanl.gov } 
\altaffiltext{2}{NASA Ames Research Center, MS 245-3, Moffett Field, CA 94035-1000; mark.s.marley@nasa.gov}
\altaffiltext{3}{Steward Observatory, University of Arizona, 933 North Cherry Ave, Tucson, AZ 85721; mcushing@as.arizona.edu}
\altaffiltext{4}{{\it Spitzer Fellow}}
\altaffiltext{5}{Joint Astronomy Center, University Park, Hilo, HI 96720; s.leggett@jach.hawaii.edu}
\altaffiltext{6}{NASA Ames Research Center, MS 245-6, Moffett Field, CA 94035-1000; thomas.l.roellig@nasa.gov}
\altaffiltext{7}{Dept. of Earth \& Planetary Sciences, Washington University, St-Louis, MO 63130-4899; lodders@levee.wustl.edu}
\altaffiltext{8}{SETI Institute, 515 North Wisham Rd., Mountain View, CA 94043; freedman@darkstar.arc.nasa.gov}

\begin{abstract}

We present the first analysis of an optical to mid-infrared spectrum of the T7.5 dwarf Gliese 570D
with model atmospheres, synthetic spectra, and brown dwarf evolution sequences.
We obtain precise values for the basic parameters of Gl 570D: 
$\teff=800$ -- 820$\,$K, $\log g\,({\rm cm/s}^2) = 
5.09$ -- 5.23, and $\log L/L_\odot= -5.525$ to $-5.551$.  The {\it Spitzer} IRS 
spectrum shows  prominent features of ammonia (NH$_3$) that can only be fitted by reducing
the abundance of NH$_3$ by about one order of magnitude
from the value obtained with chemical equilibrium models.  We model departures from 
chemical equilibrium  in the atmosphere of Gl 570D by considering the kinetics of nitrogen and 
carbon chemistry in the presence of vertical mixing.  The resulting model spectrum reproduces the data
very well.

\end{abstract}



\section{Introduction}

Ammonia is the only significant
nitrogen-bearing compound that can be readily observed in brown dwarfs.  
Its presence in very cool brown dwarf atmospheres has been expected
for some time \citep{tsu64, marley96, fl96, saumon00, all01, lf02, burrows03, saumon03}.
\citet{saumon00} reported a detection of weak NH$_3$ features in the near
infrared spectrum of the T7p dwarf Gl 229B but a clear detection had to await
the {\it Spitzer} InfraRed Spectrograph (IRS) observation of the T1/T6 binary 
$\epsilon$ Indi Bab \citep{roellig04}.  IRS spectra also reveal the presence of NH$_3$ in all observed dwarfs 
with spectral types of T2 and later \citep{cushing06}.

In addition to its importance as a strong absorber in very cool brown dwarfs,
NH$_3$ provides an important window into the phenomenon of non-equilibrium chemistry.  Because 
some chemical reactions proceed very slowly at the relatively low temperatures encountered in the
atmospheres of T dwarfs, vertical transport can lead to chemical abundances that depart
from local thermodynamic equilibrium.  This phenomenon has long been established in
Jovian planets \citep{bl78, fp85, fl94, lf94}.  In brown dwarfs, the strongest expected signatures
of departures from equilibrium chemistry are an overabundance of CO and depletion of  NH$_3$.
The prediction of  non-equilibrium chemistry in brown dwarfs \citep{fl96}
was quickly confirmed by the detection of the 4.7$\,\mu$m band of CO in Gl 229B 
\citep{noll97}, and subsequently by the detection of NH$_3$ in a high-resolution 
near-IR spectrum \citep{saumon00}.  Nonequilibrium chemistry may be common in cool
brown dwarfs since $M^\prime$ photometry shows a systematic
flux depletion in T6 -- T9 dwarfs 
that can be attributed to enhanced CO 4.7$\,\mu$m absorption \citep{golim04}
relative to models in chemical equilibrium that 

With a spectral type of T7.5 \citep{burgasser06}, Gl 570D is one of the latest T dwarfs known and its
IRS spectrum shows the strongest 
10 -- 11$\,\mu$m NH$_3$ absorption observed so far \citep{cushing06}. It is a particularly 
interesting target since it is a wide companion to 
the well-studied K4 V star Gl 570A and to a pair of M dwarfs, Gl 570BC.  The distance, metallicity, and
age of the system very effectively reduce the parameter space allowed to 
model Gl 570D. 
\citet{geballe01} used these constraints to determine the effective temperature $\teff$ and
gravity $g$ of Gl 570D and to analyze its optical and near infrared spectrum.
Here, we take advantage of the new spectroscopic information provided by the IRS
spectrum to redetermine $(\teff,g)$.  We also perform the first quantitative analysis of the
NH$_3$ 10 -- 11$\,\mu$m band for any brown dwarf.  

\section{Observations and Data Reduction}

Our analysis of Gl 570D is based on spectroscopic data covering most of its 
spectral energy distribution. We combine a Keck LRIS 0.63 -- 1.01$\,\mu$m
optical spectrum \citep{burgasser03} with a UKIRT CGS4 near IR spectrum 
covering 0.79 - 1.35$\,\mu$m and 1.43 -- 2.52$\,\mu$m \citep{geballe01}, 
and a IRS Short Low module spectrum which covers 5.43 -- 14.68$\,\mu$m 
\citep{cushing06}.
Absolute flux calibration for the UKIRT and IRS spectra was obtained by using MKO  
photometry \citep{leggett02} and IRAC band 4 photometry (B. Patten,  
priv. comm.), respectively.  The far red optical spectrum was then  
scaled to match the flux density level of the flux-calibrated UKIRT  spectrum.

\section{Models: evolution and spectra}

We analyze the spectrum of Gl 570D with a combination of model atmospheres, synthetic spectra
and evolution sequences.  Its T7.5 spectral type strongly suggests that its atmosphere is
free of condensates and we use cloudless model atmospheres
and synthetic spectra which were most recently described in \citet{fortney06} and references
therein. 
Averaging recent metallicity determinations for Gl 570A \citep{tf00, santos05, valf05} gives
[Fe/H]=0.09$\pm0.04$.  This slight enrichment is negligible in view of the other uncertainties in the
data and models and our entire analysis assumes solar metallicity.

We have developed an evolution code for brown dwarfs 
that assumes an adiabatic internal structure.  It uses
the H/He equation of state of \citet{scvh}.  Nuclear energy generation through the first branch
of the pp-chain \citep{bl93} is included.  We use the NACRE nuclear reaction rates \citep{nacre} and
ion-ion and electron-ion screening factors (Chabrier, priv. comm.)  The surface 
boundary condition is obtained from
our atmosphere models.  All brown dwarf atmosphere models become convective at depth, where the
specific entropy of the gas becomes essentially constant. The value of the entropy at the
bottom of the atmospheres $S_{\rm atm}$ gives the entropy for the adiabatic interior.
The boundary condition is described as a tabulated function
$S_{\rm atm}(\teff,g,[{\rm Fe/H}])$.  Thus, our cooling sequences are fully consistent with
the atmosphere models \citep{cb97}. 
Details of the evolution calculation will be provided in a future publication.

\section{Determination of the physical parameters}

We follow the method developed by \citet{saumon00} and initially applied to Gl 570D by 
\citet{geballe01}.  Briefly, we used the optical, near-IR and mid-IR spectra
to determine the integrated flux observed at Earth ($1.875 \pm 0.107 \times
10^{-12}\,$erg/s/cm$^2$).\footnote{The uncertainty estimate is obtained  from
the uncertainties in the absolute calibration of each piece of the SED (3\% 
for the optical spectrum, 5\% for the near-IR spectrum and 6.5\% for the IRS
spectrum).  A total uncertainty of $\pm$5.7\% is based on conservatively 
assuming that the individual calibration uncertainties all add in the same direction.
The 1\% uncertainty on the distance \citep{perryman97} is negligible by comparison.}
This represents about 70\% of the bolometric flux of Gl 570D.  We use our
grid of model spectra to compute the bolometric correction and the bolometric luminosity,
$L^{\rm s}_{\rm bol}(\teff,g)$,
which depends on the yet-to-be-determined $\teff$ and $g$.
An independent relation  $L^{\rm e}_{\rm bol}(\teff,g)$ is
given by our evolution sequences.  A consistent solution between the spectrum
and the evolution is obtained by imposing $L^{\rm s}_{\rm bol}=L^{\rm e}_{\rm bol}$.
This gives a family of solutions $\teff(g)$ (Fig. 1).  A combination of age indicators shows   
that the Gl 570 system is between 2 and 5$\,$Gyr old \citep{geballe01}, which
constrains the range of the gravity to that shown by the isochrones in Fig. 1 and the
models in Table 1.  Models A and C bracket the age limits and model B corresponds to the midpoint.
This new determination of the physical parameters of Gl 570D differs 
from that of \citet{geballe01} in several ways.  The observed SED has been
greatly extended by the {\it Spitzer} IRS spectrum and the near-IR spectrum has
been recalibrated using the latest MKO photometry.
The atmosphere models have steadily improved, most notably with the inclusion
of a more extensive line list for CH$_4$ \citep{champion92,freedl06}. Finally, the evolution is computed
with an independent code, using our cloudless atmosphere grid as the surface boundary condition,
rather than relying on the sequences of \citet{burrows97}.  Despite all these changes, the parameters
obtained for Gl 570D are remarkably similar to those of \citet{geballe01} (crosses on Fig. 1).
Most of the difference can be attributed to a shift of the isochrones in the
$(\teff,g)$ plane between our evolution sequences and those of \citet{burrows97} caused
mainly by the different surface boundary conditions.
This shows that the method is quite robust and that further improvements in the models
will not affect the values in Table 1 significantly.

Figure 1 shows that the main uncertainty in the $(\teff,g)$ of Gl 570D arises from the
uncertainly in the age of the system.  By comparison, the uncertainty
on $L_{\rm bol}$ from the flux calibration of the data is small.

Given $\teff$, $g$, the radius $R(\teff,g)$ from the evolution, and the parallax of Gl 570A,
we can compute model fluxes at Earth that can be compared directly to the data.
The model B spectrum is shown in Fig. 2.  This is not a fit of the
spectrum {\it per se}, but it is obtained solely from the procedure outlined above. 
The agreement with the optical and near-IR data is generally excellent.  The deviations
in the 1.6 -- 1.7$\,\mu$m region are due to the incomplete CH$_4$ line list.  The reason for the
underestimated $K$ band flux is unclear.  Models computed with the slightly higher metallicity 
of Gl 570A would give a higher $K$ band flux.
A similar effect could also be obtained by flux redistribution from a stronger CH$_4$ 1.6$\,\mu$m
band computed from a more complete line list.  We find that a cloudy model with a thin cloud
deck parameterized by a sedimentation
parameter $f_{\rm sed}=4$ \citep{fortney06,ackmar01}, can also improve 
substantially the agreement with the 
observed $K$ band peak.

In the mid-IR, the model systematically
underestimates the flux beyond 9$\,\mu$m, which coincides with strong absorption by NH$_3$.
The outlier models (A \& C in Table 1) give very similar synthetic spectra.
The differences remain below $\pm 0.03\,$mJy in the near-IR and below
$\pm0.2\,$mJy in the mid-IR, with higher gravity models giving lower fluxes.

\section{Nonequilibrium chemistry of NH$_3$}
 
Vertical transport in the atmosphere of a T dwarf will result in a reduced NH$_3$ abundance,
which will increase the model flux for $\lambda \wig> 9\,\mu$m and improve the agreement with the 
IRS spectrum.  We consider atmosphere models
that depart from chemical equilibrium to optimize the fit to the data.

The kinetics of the chemistry of carbon and nitrogen in giant planets and
brown dwarfs has been described in
details in \citet{lp80, fl94, gy99, lf02, bezard02, lf06} 
and the effect on the spectra of brown dwarfs in \citet{saumon03}. We only provide a brief
summary of nonequilibrium chemistry here.  

\subsection{Nonequilibrium carbon and nitrogen chemistry}

The chemistry of carbon and nitrogen in brown dwarfs atmospheres is essentially described
by the net reactions
\begin{equation}
{\rm CO} + 3 {\rm H}_2 \leftrightarrow {\rm CH}_4 + {\rm H}_2{\rm O}
\end{equation}
\begin{equation}
{\rm N}_2 + 3{\rm H}_2 \leftrightarrow 2 {\rm NH}_3.
\end{equation}
Because of the large binding energies of CO and N$_2$, both reactions proceed much more 
slowly to the right than to the left. Vertical mixing dredges up hot gas that is relatively
rich in CO and N$_2$ to the cooler part of the atmosphere, where CO and N$_2$ are converted 
very slowly into CH$_4$ and NH$_3$, respectively.  
The net result is a relative overabundance of CO and N$_2$
in the upper atmosphere and decreased abundances of CH$_4$ and NH$_3$ in the upper atmosphere.  
The abundance of
H$_2$O is also reduced by conservation of the total number of oxygen atoms.

The correct chemical pathway and reaction time scale for the conversion of CO into CH$_4$ 
remains uncertain \citep{yung88, gy99, bezard02, lf02, vf05}.
We adopt the ``fast'' chemical time scale of \citet{yung88}. For the N$_2$ conversion, we 
adopt the time scale given in \citet{lf02}.

\subsection{Vertical Transport}

Slow mixing can occur in the radiative zones of atmospheres
through a variety of processes such as the turbulent decay of waves propagated upward
from the convective/radiative boundary and instabilities arising from rapid rotation.
In the absence of any detailed modeling of these processes, we adopt a simple
model of vertical mixing by eddy turbulence \citep{gy99}.  This process is 
analogous to diffusion and occurs over a characteristic time scale
$\tau_{\rm\sss mix} \sim {H^2/K}$,
where $H$ is the pressure scale height and $K$ is the coefficient of eddy diffusion.
$K$ ranges from $\sim 10^2$ to $10^5\,$cm$^2$/s in planetary stratospheres and it is
the only free parameter in our non-equilibrium models.

The time scale for mixing in the atmospheric convection zone is
$\tau_{\rm\sss conv} \sim {H_c/v_c}$
where $H_c$ and $v_c$ are the convective mixing length and velocity, respectively,
and are evaluated with the mixing length theory.  For convenience, we choose $H_c=H$.
Note that $\tau_{\rm\sss conv} << \tau_{\rm\sss mix}$.

\subsection{Quenching of the Chemistry}

The chemical abundance profiles \citep{lf02} in 
model atmosphere B are shown in Fig. 3.  Depth in the atmosphere is indicated
by the local temperature.  The equilibrium abundances are shown with dashed lines that 
overlap solid lines for H$_2$O and CH$_4$.
The transition in the carbon chemistry from CO deep in the atmosphere 
to CH$_4$ near the surface occurs at $\log T \sim 3.25$.  
The N$_2$ to NH$_3$ transition occurs around $\log T \sim 2.9$.  Remarkably, and
in contrast to the carbon chemistry,
the N$_2$/NH$_3$ equilibrium ratio remains nearly constant deep in the atmosphere due to a near cancellation 
of the increased temperature (favoring N$_2$) and the increased pressure (favoring NH$_3$) along this
portion of the atmosphere profile.

The chemical time scales  for the conversion of CO and N$_2$ are shown by the heavy 
solid and dotted lines, respectively.  These vary over more than 20 orders of magnitude 
throughout the atmosphere and become extremely long in the upper atmosphere.
Mixing occurs rapidly in the convection zone but it is about 7 orders of magnitude slower in the 
radiative zone for our choice of $K=100\,$cm$^2$/s.  In regions where 
$\tau_{\rm\sss chem} < \tau_{\rm\sss mix}$, chemical equilibrium prevails. Where 
$\tau_{\rm\sss chem} > \tau_{\rm\sss mix}$, however, abundances of CO, CH$_4$, N$_2$, NH$_3$
and H$_2$O will depart from equilibrium (solid lines). The nonequilibrium mole fractions are approximately
determined by their values where $\tau_{\rm\sss chem} = \tau_{\rm\sss mix}$.
In our calculation,
we adopt the scheme of \citet{smith98} to determine this ``quenching level.''
As can be seen in Fig. 3, there can be more than one level
where $\tau_{\rm\sss chem} = \tau_{\rm\sss mix}$, resulting in alternating zones in and out
of chemical equilibrium.  For simplicity, we consider only the uppermost (lowest $T$) crossing.
It turns out that in all cases, this is an excellent approximation to the full
solution for the chemical profile since the deeper quenching occurs well below the level of formation 
of the corresponding molecular bands.

Figure 3 shows that above the CO quenching level
($\log T \sim 3.03$) the nonequilibrium abundance of CO is orders of magnitude higher than at 
equilibrium.  The abundances of CH$_4$ and H$_2$O are barely affected for this choice of $K$
(slow mixing).
The NH$_3$ abundance is reduced by about one order of magnitude in the upper 
atmosphere.  
In this model, N$_2$ is quenched below the bottom of the atmosphere model, in a region where
N$_2$/NH$_3$ is still nearly constant.  The fact that the equilibrium mole fraction of NH$_3$
is nearly constant for $\log T \wig>3.1$ implies that the non-equilibrium NH$_3$ abundance
profile is completely insensitive to the choice of the eddy mixing coefficient $K$ over a very
wide range ($\log K \wig> -8$).  This means that {\it all} plausible choices of $K$ will
give the same NH$_3$ abundance and therefore the same mid-IR spectrum. 

\subsection{Optimal model}

In view of the above remarks, we arbitrarily choose $K=100$\,cm$^2$/s to compute
nonequilibrium spectra for the three models in Table 1. Besides the choice of equilibrium or nonequilibrium
chemistry, the only fitting parameter left is the gravity, constrained to be between the values for
models A and C (Table 1).  We determine the goodness of fit of a 
synthetic spectrum by computing the $\chi^2$
between the model and the data for $\lambda \ge 9\,\mu$m, which is where NH$_3$ features are
seen in the spectrum (Fig. 2).  For this purpose, model spectra are smoothed to the IRS Short Low
point source spectral resolution of $\lambda/\Delta\lambda=126$ with a Gaussian filter and then
binned to the instrument's wavelength sampling. The uncertainty ($\sigma$) on the fitted 
$\chi^2$ is obtained by fitting 
5000 simulated data sets generated by adding a random Gaussian noise distribution to each 
pixel with a width given by the individual pixel noise ($\sim 0.1\,$mJy).  We also fit renormalized 
spectra to account for the $\pm6.5$\% calibration uncertainty, which turns out to be the dominant 
source of uncertainty.  Nonequilibrium model C always gives the best fit.  A reasonable fit
($2.8\sigma$ worse than model C) is
obtained with nonequilibrium model B if the observed flux is increased by the full 6.5\% uncertainty.
The best equilibrium model fit is $6.7\sigma$ worse than the best nonequilibrium model fit.

The nonequilibrium model C spectrum is shown by the dashed (blue) curve on 
Fig. 2.  The agreement with the data is remarkable considering that all we have done is
include non-equilibrium chemistry to the calculation and increase the gravity slightly, with no other
free parameter.
We conclude that the NH$_3$ abundance is reduced by nonequilibrium chemistry and that the parameters of
Gl 570D fall between models B and C, and much more likely closer to the latter.

\section{Alternative explanations for the NH$_3$ depletion}

We briefly consider five possible explanations that could account for the enhanced
mid-IR flux in Gl 570D: a low nitrogen abundance, an error in the temperature profile
of the atmosphere, uncertain NH$_3$ opacity, time variability, and
photodissociation of NH$_3$.

\subsection{Low nitrogen abundance}

The metallicity of Gl 570D is very nearly solar, based on abundance determinations of
Na, Si, Ca, Ti, V, Cr, Fe, Co, and Ni in Gl 570A that range from [X/H]$=-0.10$ to 0.24, 
with a typical spread of
$\sim 0.1$ between independent determinations \citep{santos05, tf00, valf05}.  To the 
best of our knowledge however, the N abundance has not been measured.
A low [N/H] value would result in a decreased NH$_3$ abundance that could potentially 
explain the relatively high mid-IR flux.  The dependance of the
mole fraction of ammonia on the nitrogen abundance is $\log X_{{\rm NH}_3} \sim 0.5$[N/H]
\citep{lf02}.  In our models, the
strongest NH$_3$ absorption occurs at a depth of $\log T=2.65$ in the atmosphere 
where the non-equilibrium chemistry reduces $X_{{\rm NH}_3}$ 
by a factor of 10 (Fig. 3).  To obtain a similar reduction from a low nitrogen abundance would
require [N/H]$=-2$, which would be an extreme departure from the abundances of
other metals in Gl 570A.

In principle, N-bearing condensates may also cause a depletion of N-bearing gases in the 
atmosphere. However, there are no N-bearing condensates expected along the 
$(T,P)$ profile of Gl 570D that can significantly remove nitrogen, and temperatures 
are not low enough to allow 
NH$_4$SH or NH$_3$ condensation as in the much cooler upper atmospheres of the Jovian planets.

\subsection{Uncertainties in the temperature profile of the atmosphere}

Our determination of $L_{\rm bol}$ and our solution for $(\teff,g)$ are in excellent agreement
with the values obtained by \citet{geballe01} without the benefit of the IRS spectrum. This indicates that
while the near-IR flux has remained constant, the unknown mid-IR flux of Gl 570D in March 2000 could 
not be significantly 
different from that measured in Feb 2005.  The fact that our equilibrium model is in very good agreement
with the full SED of Gl 570D (Fig. 2) supports the notion that all pieces of the SED are consistent with
each other.   Finally, $\teff$  would have to increase to $\sim 1050\,$K to obtain an equilibrium NH$_3$
mole fraction similar to the non-equilibrium value needed to fit the IRS spectrum.
Such a high $\teff$ is completely incompatible with {\it all} of the Gl 570D observations. 
If we consider the temperature profile of the atmosphere, an increase of over 200$\,$K ($\sim40$\%) at 
$P \wig< 1\,$bar would be required; an implausible error in our models, an inconsitent with the near-IR
spectrum of Gl 570D. 

\subsection{Uncertainties in the NH$_3$ opacity}

The opacity of NH$_3$ is computed from a line list obtained by combining the GEISA \citep{husson92} 
and HITRAN \citep{rothman05} data bases, complemented with recent laboratory measurements
and theoretical calculations (see \citet{burrows97} for more details).  To our knowledge, this compilation of NH$_3$
opacity is the most complete presently available.  Nonetheless, the line list is only complete
up to $T=300\,$K and transitions arising from levels that are excited only at higher temperatures are
underrepresented or missing.  Qualitatively, the NH$_3$ opacity we calculate for Gl 570D is a lower 
limit to the actual opacity, implying that the derived underabundance of NH$_3$ is an {\it upper} limit.
More quantitatively, the 9--14$\,\mu$m band of NH$_3$ arises from the ground state
($\nu_0 \rightarrow \nu_2$), with a small overlaping contribution from the $\nu_2$ excited state 
($\nu_2 \rightarrow 2\nu_2$).  The energy differences between other excited vibrational 
states of NH$_3$ is too large for the corresponding hot bands to overlap with the $\nu_0 \rightarrow \nu_2$ band.
We are confident that the line list for the  9--14$\,\mu$m band on which our 
analysis of the abundance of NH$_3$ is based is quite complete.

\subsection{Time variability of Gl 570D}

The spectroscopic data analyzed here was gathered over a period of several years:
The optical spectrum is from observations taken in March 2000 and Feb 2001 \citep{burgasser03}, the near-IR 
spectrum in March 2000 \citep{geballe01} and the {\it Spitzer} IRS spectrum in
February 2005 \citep{cushing06}.  It is conceivable that the luminosity of Gl 570D may have varied
during this 5 year period, affecting our determination of $\teff$ and $g$ and fortuitously
result in our excellent fit of the combined spectrum with a model departing from chemical equilibrium.  
\citet{burgasser03} do not report any variation in the optical flux of Gl 570D between their
two observations taken one year apart, and the reported equivalent width of Cs I and Rb I lines
show no variation within 1.5$\sigma$.  The near IR spectrum is calibrated to $JHK$ MKO photometry
taken in Feb 2000 \citep{geballe01}.  Repeat observations in July 2005 are fully consistent ($<1\sigma$) with the
original photometry.  The 2MASS photometry of Gl 570D was obtained in May 1998.  Applying the
color transformations of \citet{sl04} shows agreement to $2\sigma$ ($J$), $1\sigma$ ($H$) and
$<1\sigma$ ($K$) with the MKO magnitudes.  Based primarily on the stability of the near-IR MKO photometry, we 
conclude that any time variability in Gl 570D
over a period of 7 years is at most at the 2-3\%  level.  This is less than our estimate of a 5.7\%
uncertainty on $L_{\rm bol}$.

Nevertheless, we consider the possibility that the absolute mid-IR flux may not be representative of the
average or typical state of Gl 570D.  If we relax the absolute flux calibration of the IRS spectrum, 
do we still see a NH$_3$ depletion?  For this purpose, we have repeated the fitting procedure
described in \S 5.4, with the flux calibration adjusted by a factor that minimizes the value of
$\chi^2$. Allowing for the estimated noise in each pixel of the observed spectrum, we find that the
non-equilibrium models always fit better than the equilibrium models at the $\sim 17\sigma$ level,
with insignificant variation between models A, B and C.

\subsection{Photodissociation of NH$_3$}

NH$_3$ is a molecule that is easily dissociated by UV photons $\sim 1900\,$\AA.  Since Gl 570D is a companion
to a main sequence star, it is possible that the UV flux from the primary results in a depletion of
NH$_3$ in Gl 570D.  This process was discussed in details by \citet{saumon00} in the case of another
late T dwarf with a companion, Gl 229B, where it was found to be completely negligible at the level of
the photosphere.  By comparison, the UV flux of the dK4 star Gl 570A is about 10 times larger than that of 
the dM1 star Gl 229A (based on the NextGen models of \citet{ah99}) but the flux incident on Gl 570D is
smaller because of the much greater projected separation
(1525$\,$AU \citep{burgasser00} rather than 44$\,$AU \citep{golim98} for Gl 229B).  Since the atmospheres of both 
brown dwarfs are quite similar, we can
expect that the photodissociation of NH$_3$ in Gl 570D will be $10\times(44/1525)^2 \sim 0.008$ times
that in Gl 229B.  The Gl 570 BC pair of M dwarfs lies about 10 times closer to Gl 570A than Gl 570D
and their contribution to the UV flux incident on Gl 570D is a small correction to the above estimate.

\section{Conclusion}

Our method to determine $\teff$ and the gravity of brown dwarfs is most effective when
applied to objects with known parallax and which are members of systems where the
primary provides the metallicity and age indicators.  The method is based on integrated fluxes
and evolution sequences rather than a direct fit of the spectrum, an approach that
largely eliminates biases due to remaining systematic problems in the model spectra.  With 
recalibrated optical and near-IR spectra and, most importantly, a {\it Spitzer} IRS
spectrum, we have redetermined $(\teff,g)$ for Gl 570D to find nearly the
same values as \citet{geballe01}.  

An excellent fit of the IRS spectrum can only be obtained if the NH$_3$ abundance
reduced by nearly one order of magnitude in the upper atmosphere.   This is most
readily explained by vertical mixing that drives the nitrogen chemistry away
from chemical equilibrium \citep{lf02}.  We have considered several other possible explanations
but none of them are plausible.  

We find that in late T dwarfs, the
nonequilibrium N$_2$/NH$_3$ ratio generally does not depend on the efficiency of mixing in the
atmosphere.
Therefore, while the 10 -- 11$\,\mu$m band of NH$_3$ is a sensitive indicator of nonequilibrium
chemistry in brown dwarf atmospheres, it provides little information about the 
efficiency of mixing.
On the other hand, the CO abundance is very sensitive to the choice of the mixing parameter $K$,
and photometric or spectroscopic observations of the 4.7$\,\mu$m CO band is the most 
effective way to determine $K$.
Our study provides the most accurate parameters of any known brown dwarf (Fig. 1): $\teff=800$ -- 820$\,$K,
$\log g=5.09$ -- 5.23, $R/R_\odot= 0.0901$ -- 0.0853, $\log L/L_\odot=-5.525$ to $-5.551$, 
$M/M_{\rm J}=38$ -- 47 and an age of 3 to 5Gyr.  The upper limit of this range
(Fig. 1) is at least $2.8\sigma$ more likely than the lower limit.
Gl 570D joins Gl 229B as the second T dwarf where the effect of nonequilibrium chemistry 
due to vertical transport has been demonstrated explicitly.

This work is based (in part) on observations made with the Spitzer Space
Telescope, which is operated by the Jet Propulsion Laboratory,
California Institute of Technology under NASA contract 1407.
T. Roellig and M. Marley acknowledge the support of the NASA Office of
Space Sciences. K. Lodders is supported by NSF grant AST04-06963. Part of this work was
supported by the United States Department of Energy under contract W-7405-ENG-36, and NASA through the 
Spitzer Space Telescope Fellowship Program, through a contract issued by the Jet Propulsion 
Laboratory, California Institute of Technology under a contract with NASA. 
This research has benefited from the M, L, and T dwarf compendium housed at 
DwarfArchives.org and maintained by C.~R. Gelino, J.~D. Kirkpatrick, and A.~J. Burgasser.

\clearpage

\begin{figure}
\plotone{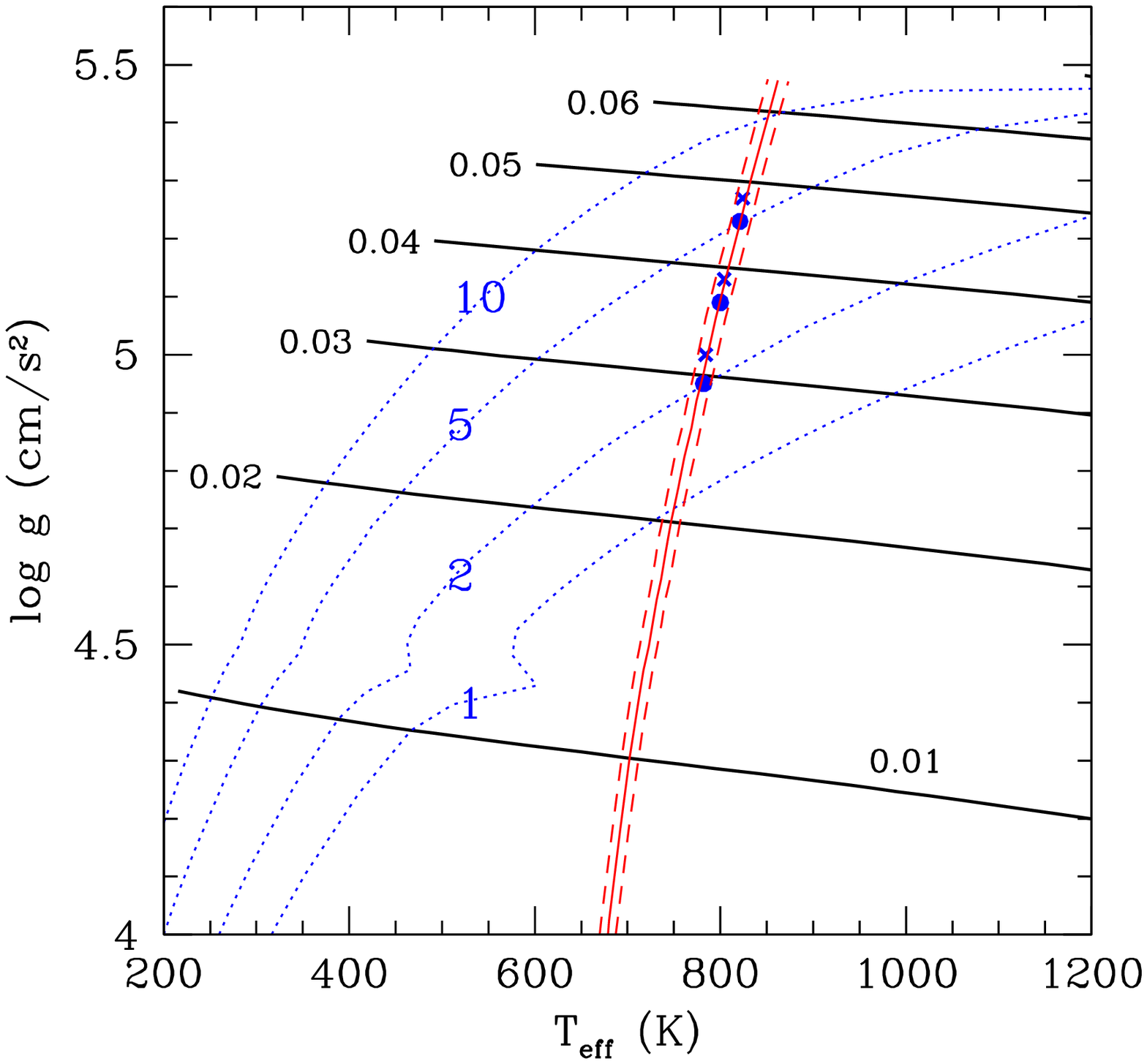}
\caption{$\teff$ and gravity for Gl 570D.  Our brown dwarf evolution
tracks are shown by heavy black lines labeled with the mass in $M_\odot$.
Isochrones (blue dotted lines) are labeled in Gyr.  
The nearly vertical lines (solid and dashed, red) are the locus of $(\teff,g)$ points with
$\log L/L_\odot = -5.525 \pm 0.024$.  The solid dots show the three model of Table 1.
For comparison, the crosses show the values obtained by \citet{geballe01}.
[{\it See the electronic edition of the Journal for a color version of this figure.}]}
\label{fig:fig1}
\end{figure}

\clearpage

\begin{figure}
\plotone{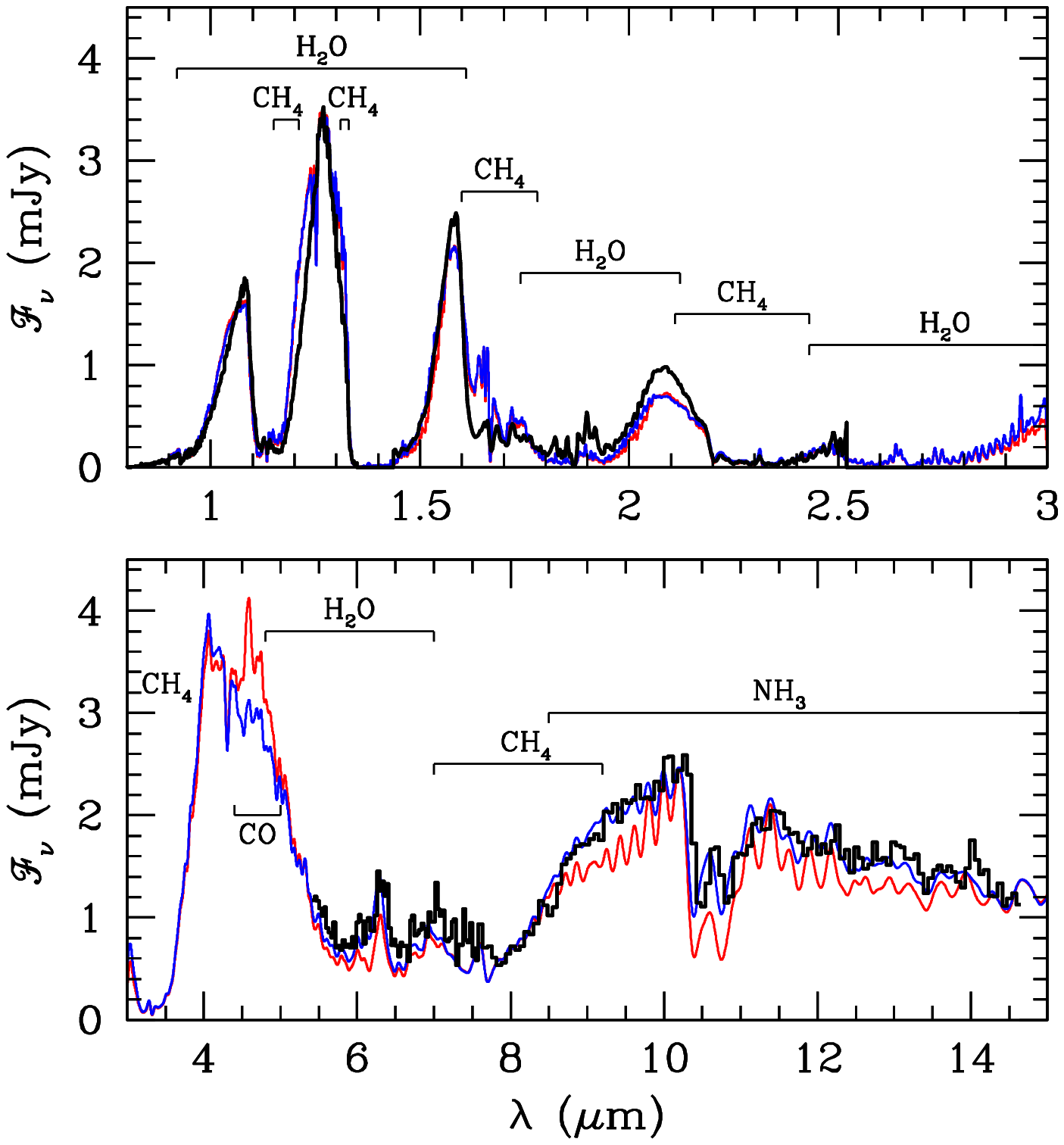}
  \caption{Comparison of the observed and modeled SED of Gl 570D.  The 
           data (heavy black curve and histogram) are from \citet{burgasser03},
           \citet{geballe01} and \citet{cushing06}.
           The equilibrium model B (Table 1)  is shown by the thin (red) solid line,
           and the best fitting nonequilibrium model (model C with $\log K=2$) 
           is shown by the dashed (blue) line.
           The main molecular absorbers are indicated.  The models are plotted at a
           spectral resolution of R=500 (upper panel) and R=100 (lower panel), approximating
           the resolution of the data.  
[{\it See the electronic edition of the Journal for a color version of this figure.}]}
\label{fig:fig2}
\end{figure}

\clearpage

\begin{figure}
\plotone{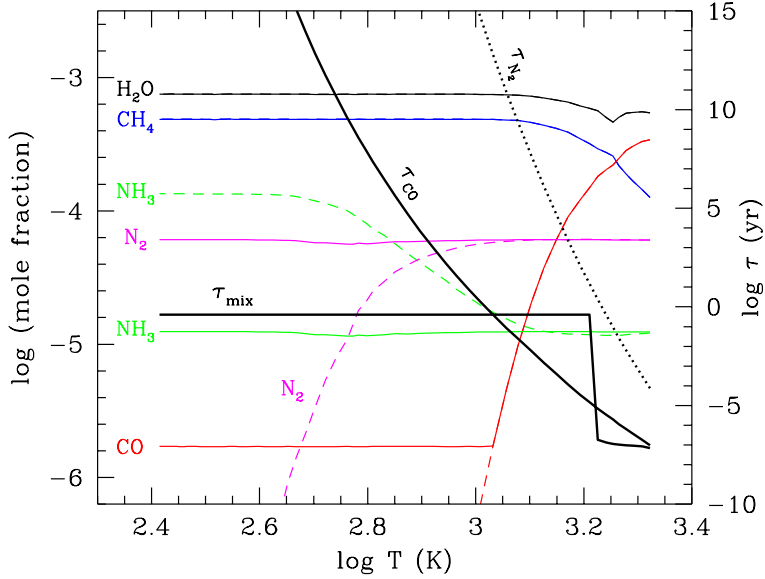}
  \caption{Chemical profile of model atmosphere B (Table 1).  Mole fractions of
           H$_2$O, CH$_4$, N$_2$, NH$_3$ and CO are shown in equilibrium (dashed curves)
           and out of equilibrium ($\log K=2$, solid curves) as a function of
           the temperature in the model atmosphere.  The pressure
           increases from the top of the atmosphere (left) toward the bottom (right).
           Heavy black lines show the mixing time scale ($\tau_{\sss mix}$) and the
           time scale for the destruction of CO ($\tau_{\sss CO}$, solid) and 
           N$_2$ ($\tau_{\sss N_2}$, dotted).  Time scales are given on the right axis.  
           The mixing time scale is nearly discontinuous where the atmosphere becomes 
           convective ($\log T \wig> 3.2$).  The 10--11$\,\mu$m NH$_3$ band is formed in the
           $\log T=2.65$ to 2.9 region.
[{\it See the electronic edition of the Journal for a color version of this figure.}]}
\label{fig:fig3}
\end{figure}

\clearpage

\tablecolumns{7}
\begin{deluxetable}{ccccccc}
\tablewidth{0pt}
\tablecaption{Range of physical parameters of Gliese 570D}
\tablehead{
\colhead{Model} & \colhead{$\teff$}  &  \colhead{$\log g$}  & \colhead{$\log L/L_\odot$}
 & \colhead{Mass} & \colhead{Radius} & \colhead{Age} \\
\colhead{}  & \colhead{(K)}  & \colhead{(cm/s$^2$)} &   \colhead{}  & \colhead{$(M_J)$}  & \colhead{$(R_\odot)$} & \colhead{(Gyr)}}
\startdata
 A & 782 & 4.95 & $-5.503$ & 31 & 0.0950 &\phs 2.0 \phs   \\
 B & 800 & 5.09 & $-5.525$ & 38 & 0.0901 &\phs 3.2 \phn   \\
 C & 821 & 5.23 & $-5.551$ & 47 & 0.0853 &\phs 5.0 \phn   \\
\enddata
\end{deluxetable}

\end{document}